\documentclass[aps,prl,twocolumn,groupedaddress]{revtex4}

\usepackage{epsf}
\usepackage{graphicx}
\usepackage{dcolumn}
\usepackage{bm}

\begin{document}

\title{Nonlinear resistance of 2D electrons in crossed electric and magnetic fields.}

\author{Jing Qiao Zhang}
\author{Sergey Vitkalov}
\email[Corresponding author: ]{vitkalov@sci.ccny.cuny.edu}
\affiliation{Physics Department, City College of the City University of New York, New York 10031, USA}
\author {A. A. Bykov}
\affiliation{Institute of Semiconductor Physics, 630090 Novosibirsk, Russia}

\date{\today}

\begin{abstract} 
The longitudinal resistivity of two dimensional (2D) electrons placed in strong magnetic field is significantly reduced by applied electric field, an effect which is studied in a broad range of magnetic fields $B$ and temperatures $T$ in GaAs quantum wells with high electron density. The data are found to be in good agreement with theory, considering the strong nonlinearity of the resistivity as result of non-uniform spectral diffusion of the 2D electrons. Inelastic processes limit the diffusion. Comparison with the theory yields the inelastic scattering time $\tau_{in}$ of the two dimensional electrons. In the temperature range $T=2-10K$ for overlapping Landau levels, the inelastic scattering rate $1/\tau_{in}$ is found to be proportional to $T^2$, indicating a dominant contribution of the electron-electron scattering to the inelastic electron relaxation.  In a strong magnetic field, the nonlinear resistivity demonstrates scaling behavior, indicating a specific regime of electron heating of well-separated Landau levels. In this regime the inelastic scattering rate is found to be proportional to $T^3$, suggesting the electron-phonon scattering as the dominant mechanism of the inelastic relaxation. At low temperatures and separated Landau levels an additional regime of the inelastic electron relaxation is observed: $\tau_{in} \sim T^{-1.26}$.       

\end{abstract}
  

\maketitle

\section{Introduction}

The nonlinear properties of low-dimensional electron systems attract a great deal of attention for its fundamental significance as well as for potentially important applications in nanoelectronics. In response to microwave radiation and $dc$ bias, strongly nonlinear electron transport\cite{zudov2001,engel2001,yang2002,dorozh2003,willett2004,mani2004,kukushkin2004,stud2005,bykov2005,bykovJETP2006,bykov2007R,zudov2007R,du2007,stud2007,zudovPRB2008,gusev2008a,hatke2009a,hatke2009b,dorozh2009,durst2003,ryzhii1970,anderson,shi,liu2005,dietel2005,inarreaPRB2005,vavilov2004,dmitriev2005,alicea2005,volkov2007,glazman2007,dmitriev2007} that gives rise to unusual electron states \cite{mani2002,zudov2003,zudov2007,bykov2007zdr,zudov2008zdr,andreev2003,auerbach2005} has been reported in two-dimensional systems of highly mobile electrons in a high magnetic field. There has also been great interest in the nonlinear response of quantum ballistic constrictions, where the effects of quantum interference, spatial dispersion and electron-electron interaction play essential roles \cite{dicarlo,wei,leturcq,zumbhl,lofgren,zhang2006,brouwer,vavilovnl,sanchez,spivak,polianski,andreev2006}.

Recent experiments, in which a $dc$ electric field applied to highly mobile 2D electrons placed in strong magnetic fields, have demonstrated a variety of fascinating nonlinear phenomena \cite{yang2002,bykov2005,bykov2007R,zudov2007R,gusev2008b,zudov2009}. Oscillations of the nonlinear magnetoresistance with a magnetic field, which appear at a finite $dc$ bias, have been reported \cite{yang2002,bykov2005,bykov2007R,zudov2007R}. These interesting oscillations, decaying at high temperatures \cite{zudov2009},  are attributed to Landau-Zener transitions between Landau levels \cite{yang2002}. At substantially smaller $dc$ biases another important class of nonlinearities has been identified \cite{bykov2007R,gusev2008b}. 

In this paper we study in detail the effect of the small $dc$ electric field $E$ on the longitudinal resistance of two-dimensional electrons in GaAs quantum wells placed in a strong magnetic field. In such a magnetic field the density of states of the 2D electrons is modulated due to the Landau quantization of the electron motion. The electric field $E$ decreases the resistance significantly \cite{bykov2005,bykov2007R,zudov2007R,gusev2008b}. The effect, existing in a broad range of temperatures, can not be explained by an increase of the electron temperature due to the heating by the electric field $E$ \cite{bykov2007R,romero2008warming}. In the paper \cite{bykov2007R} the effect  is attributed to a non-uniform spectral diffusion of the 2D electrons induced by the electric field \cite{dmitriev2005}. The spectral diffusion produces a specific distribution of 2D electrons in the quantized spectrum, which is significantly different from the canonical Fermi-Dirac form. In fact the observed strong nonlinearity is result of the deviations of the electron distribution from the Fermi-Dirac function. The effect is considerably enhanced in electron systems with high mobility and high electron density. The high electron mobility provides strong absolute variations of the density of states and the spectral diffusion with electron energy, increasing appreciably the magnitude of  the non-temperature deviations. The high electron density provides substantial decrease of the electron-electron scattering, which makes the relaxation of the deviations to be weak.

Effects of an electric field $E$ on the resistance of two dimensional electrons placed in  strong magnetic fields  have been studied in many works \cite{heating,pinch}.
Substantial part of these studies was focused on an effect of the electric field $E$ on an amplitude of quantum oscillations of the resistivity.  The quantum (Shubnikov de Haas, SdH) oscillations are result of the quantization of the electron spectrum in strong magnetic field \cite{shoenberg1984}.
The amplitude of the oscillations depends significantly on the electron temperature \cite{shoenberg1984,ando}. It has been found that the amplitude of the SdH oscillations decreases with the electric field $E$  \cite{heating}. The effect is attributed to an increase of the electron temperature  $ T_e$ due to the electric heating. The explanation is based on an assumption that the surplus of the Joule energy provided by the electric field $E$ is rapidly shared among the carriers through electron-electron interaction, establishing the thermal (Fermi-Dirac) distribution at an elevated temperature $T_e$ \cite{dolgopol1985,pepper}.  The $T_e$ approximation works well in systems with a strong electron-electron scattering. It ignores any deviations of the non-equilibrium electron distribution from the Fermi-Dirac form. The approximation has been widely and successfully used for 2D electron systems with low electron density and/or mobility \cite{heating}. We note, however, that a substantial discrepancy between the temperature $T_e$, obtained from the analysis of the amplitude of the quantum oscillations in the $T_e$ approximation, and the one obtained, using another experimental method, has been reported in GaAs 2D systems with a high electron mobility \cite{pepper}. 

Despite the apparent applicability of the $T_e$ approximation to the overheated electron systems, recent studies have revealed an inadequacy of the temperature description of the nonlinear transport of highly mobile 2D carriers \cite{bykov2007R,romero2008warming,gusev2008b}.  Instead of the $T_e$ approximation in this paper we use a different approach \cite{dmitriev2005}. Below we evaluate the distribution function, using an equation of the spectral diffusion. In the computations any assumptions regarding the shape of the electron distribution function are relaxed. In contrast to the $T_e$ approximation the new approach to the heating via the direct evaluation of the electron distribution function is more universal and accurate. It takes into account, in principle, $both$ the broadening ("temperature" increase) of the distribution function $and$ the deviations of the distribution function from Fermi-Dirac form in response to the electric field $E$. The later appears to be the dominant source of the strong nonlinearity observed in highly mobile 2D electron systems at small electric fields.

 The spectral diffusion is limited by an electron inelastic relaxation, which moves the electron system back to thermal equilibrium. It opens new possibilities to study inelastic processes and nonlinear electron kinetics of low dimensional systems.  In the present paper we explore these possibilities. We study the effect of electric fields on the resistivity in a broad range of magnetic fields and temperatures. We compare the experimental results with numerical simulations of the spectral diffusion. The comparison gives the inelastic scattering time of 2D electrons in a broad range of magnetic fields and temperatures. 

In the temperature interval $T=2-10K$ for overlapping Landau levels, the inelastic scattering rate $1/\tau_{in}$ is found to be proportional to the square of the temperature, indicating the dominant contribution of the electron-electron interaction into the relaxation of the electron distribution function.  At a strong magnetic field, at which Landau levels are well separated, the nonlinear resistance demonstrates an interesting scaling behavior. In this regime at high temperatures the inelastic scattering rate is found to be proportional to $T^3$, indicating leading contribution of the electron-phonon scattering to the inelastic relaxation. At low temperature and separated Landau levels an additional regime of the inelastic electron relaxation is observed: $\tau_{in} \sim T^{-1.26}$.

 The paper has the following organization. The "Experimental Setup" section presents the main kinetic parameters of samples and details of the experiment. The "Theory and Numerical Simulations" section presents basic components of the theory and discusses essential steps used to calculate the longitudinal resistance. Experimental results and a comparison with numerical simulations are presented in the section "Results and Discussion". Section "Conclusion" contains a summary of the research.

\section{Experimental Setup} 

Our samples are high-mobility GaAs quantum wells grown by molecular beam epitaxy on semi-insulating (001) GaAs substrates. The width of the GaAs quantum well is 13 nm. Two AlAs/GaAs type-II superlattices grown on both sides of the well served as barriers, providing a high mobility of 2D electrons inside the well at a high electron density\cite{fried1996}. Two samples (N1 and N2) were studied with electron density $n_1$ = 12.2 $\times 10^{15}$ m$^{-2}$, $n_2$=8.2 $\times 10^{15}$ (m$^{-2}$) and mobility $\mu_1$= 93 m$^2$/Vs, $\mu_2$=85 (m$^2$/Vs) at T=2.7K. At higher densities the cyclotron radius $r_C$ of 2D electrons at Fermi level is larger. As it is shown below, this increases the spectral diffusion and the nonlinear response in strong magnetic fields. 

\begin{figure}[tbp]
\vbox{\vspace{0 in}} \hbox{\hspace{+0.1in}} \epsfxsize 3.0 in 
\vskip -0.5in %
\epsfbox{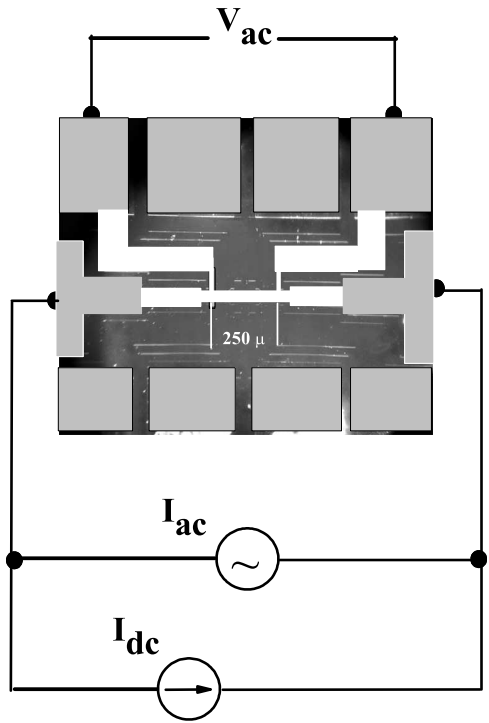} \vskip 0.5in
\caption{ Schematic view of experimental setup. Studied 2D electron system is etched in the shape of a Hall bar. White area schematically presents the  details of the Hall bar: the width and the length of the measured part of the sample are $d=$50 $\mu m$ and $L=$250 $\mu m$. Direct current $I_{dc}$ is applied simultaneously with $ac$ current $I_{ac}$ through current contacts formed in the 2D electron layer. The longitudinal $ac$ voltage $V_{ac}$ is measured between potential contacts displaced 250 $\mu m$ along each side of the sample. }
\label{sample}
\end{figure}

Measurements were carried out between T=0.3K and T=30K in a He-3 insert in a superconducting solenoid. Samples and a calibrated thermometer were mounted on a cold cooper finger in vacuum. Magnetic fields up to 1 T were applied perpendicular to the 2D electron layers patterned in a form of $d$=50 $\mu m$ wide Hall bars with a distance of 250 $\mu m$ along the bars between potential contacts.  A schematic view of  experimental setup is shown in Fig.\ref{sample}. To measure the resistance we have used the four probes method. Direct electric current $I_{dc}$ ($dc$ bias) is applied simultaneously with an $ac$ excitation $I_{ac}$ through the same current contacts (x-direction). The current contacts are placed far away from the measured area at a distance of 500 $\mu m$, which is much greater than the inelastic relaxation length of the 2D electrons $L_{in}=(D \tau_{in})^{1/2} \sim 1-5 $ $\mu m$ (see below). The later insures that possible nonlinearities near the current leads provide negligibly small contribution to the total nonlinear response measured in the experiments. 

Experiments are done at fixed magnetic fields corresponding to maximums of the Shubnikov de Haas oscillations. At this condition the Fermi level is located at a maximum of the density of states and contributions of the edge states to the total electron transport is small. Below we consider the density of the electrical current across the samples to be a constant.      

The longitudinal voltage $V_{ac}$ was measured between potential contacts (displaced along the x-direction) using a lockin amplifier with 10 M$\Omega$ input impedance. In the experiments the potential contacts provided insignificant contribution to the overall nonlinear response due to small values of the contact resistance (about 1k$\Omega$) and negligibly small electric current flowing through the contacts ($<0.1$ nA).  

The differential longitudinal resistance $r_{xx}=V_{ac}/I_{ac}$ is measured at a frequency of 77 Hz in the linear regime. In the experiment a dependence of differential resistance $r_{xx}=dV_{xx}/dI$ on the $dc$ bias $I_{dc}$ is measured. The resistance $R_{xx}$ of the sample is obtained by an integration of the differential resistance: $R_{xx}= (\int r_{xx}dI)/I_{dc}$. In the paper we compare the resistance $R_{xx}$ with numerical calculations based on recent theory \cite{dmitriev2005}.     

Experiments are done in a classically strong magnetic fields ($\omega_c \tau_{tr} \gg 1$), where the $\omega_c$ is cyclotron frequency and $\tau_{tr}$ is the transport scattering time. At this condition the electric current density $\vec J=(J_x, 0)$ directed along the x-axes is almost perpendicular to the total electric field $\vec E=(E_x,E_y)$, where $E_x \ll E_y$ \cite{ziman}. The magnitude of the Hall electric field $E_H=E_y$ directed along the y-axes is almost equal to the magnitude of the total electric field $\vert \vec E \vert$. Below we consider the magnitude of the Hall electric field $E_H$ to be equal to the magnitude of the total electric field $\vec E$ applied to the samples. The local Joule heat injected into the 2D systems per second can be evaluated with an accuracy better than 2\% as: $J_x \cdot E_x = (\sigma_{xx} E_x+\sigma_{xy} E_y)\cdot (\sigma_{xx}/\sigma_{xy})E_y \approx \sigma_{xx} \cdot E_H^2$, where $\hat \sigma$ is the conductivity in the strong magnetic field. 

In our experiments the Hall voltage $V_{xy}$ is recorded simultaneously with the longitudinal voltage $V_{xx}$.  Observed variations of the Hall conductivity $\sigma_{xy}$ and the Hall electric field $E_H$ with the $dc$ bias were below 1\%. These variations yield a negligibly small contribution to the overall dependence of the longitudinal conductivity $\sigma_{xx}$ on the $dc$ bias. This contribution are ignored in the comparison between the experiment and the theory.          

\section {Theory and Numerical Simulations}

In this section we present basic parts of the theory \cite{dmitriev2005} and details of the numerical calculations of the nonlinear resistivity. The theory considers nonlinear electron transport in a strong magnetic field. In the magnetic field the electron spectrum is quantized and the density of states oscillates with the energy. The period of the oscillations is the cyclotron energy $\hbar \omega_c$.  The width of the Landau levels is $\Gamma=\hbar/\tau_q$, where $\tau_q$ is quantum scattering time. At low temperatures the time $\tau_q$ is determined by an elastic impurity scattering of the 2D electrons. At small quantized magnetic fields the electron spin splitting is much smaller the level width $\Gamma$ \cite{romeroHparallel2008}. The spin splitting is neglected in the paper.

The net longitudinal conductivity of the 2D electrons $\sigma_{nl}=\sigma_{xx}$ is a sum of conductivities $\sigma(\epsilon)$ of the  levels with energy $\epsilon$ over all possible energies, weighted with the first derivative of the distribution function $\partial f/\partial \epsilon$ \cite{ando}:

\begin{equation}
\sigma_{nl}= \int \sigma(\epsilon)(-\partial f/ \partial \epsilon) d\epsilon,
\label{sigma_nl} 
\end{equation}

In the leading approximation for a classically strong magnetic field   the longitudinal conductivity $\sigma(\epsilon)$ at an energy $\epsilon$ reads \cite{dmitriev2005}:
 
\begin{equation}
\sigma(\epsilon)=\sigma_D \tilde{\nu}^2(\epsilon),
\label{sigma_dc} 
\end{equation}
where $\sigma_D=e^2 \nu_0 v_F^2/2 \omega_c^2 \tau_{tr}$ is the $dc$ Drude conductivity in a strong magnetic field $B$, $\tilde{\nu}(\epsilon)= \nu(\epsilon)/ \nu_0$ is dimensionless density of states (DOS),  $\tau_{tr}$ and $\nu_0=m/\pi \hbar^2$ are transport scattering time and the density of states  at zero magnetic field and $v_F$ is the Fermi velocity. The approximation neglects effects of the electric field on the electron-impurity collision, which yields a negligibly small correction to the nonlinear resistance at small electric fields\cite{dmitriev2005}. The dominant nonlinear effect is due to a non-trivial energy dependence of the distribution function $f(\epsilon)$, which is a result of non-uniform spectral diffusion of the 2D electrons in response to the total $dc$ electric field $\vec E$ applied to the system. 

Due to conservation of total electron energy $\epsilon_0$ in the presence of the external electric field $\vec E$ and the elastic electron-impurity scattering, the kinetic energy of an electron $\epsilon_K$ depends on the electron position $\vec r$: $\epsilon_ K(\vec r)=\epsilon_0 -e \vec E \vec r$. As a result of the energy conservation, the diffusion motion of the electron in real space originates a diffusion of the electron kinetic energy in the energy space. The diffusion generates a spectral electron flow from occupied electron levels below the Fermi energy to empty states above it. The coefficient of the spectral diffusion $D_\epsilon(\epsilon)$ is proportional to the coefficient of the spatial diffusion $D(\epsilon)= v_F^2 \tilde{\nu}(\epsilon) /2 \omega_c^2 \tau_{tr}= r_C^2 \tilde{\nu}(\epsilon) /2 \tau_{tr}$: $D_\epsilon(\epsilon)=(e E)^2 D(\epsilon) \sim (\delta \vec r)^2$.  The spectral diffusion is proportional to square of the cyclotron radius $r_C$ and the normalized density of states $\tilde{\nu}(\epsilon)$. The spectral diffusion is most effective in the center of the Landau levels, where the density of states is high, gradually decreases away from the center and is suppressed considerably between Landau levels, where the density of states is small. 

The spectral diffusion is described by the Fokker-Plank type equation \cite{dmitriev2005}:

\begin{equation}
-\frac{\partial f}{\partial t}+E^2\frac{\sigma _{dc}^D}{\nu _0 \tilde{\nu}(\epsilon)}\partial_{\epsilon}\left[\tilde{\nu} ^2(\epsilon) \partial _{\epsilon}f(\epsilon)\right]=\frac{f(\epsilon)-f_T(\epsilon)}{\tau_{in}}
\label{main}
\end{equation}

The left side of the equation describes the spectral diffusion of a spherical part of the electron distribution function $f$ induced by the electric field $E$ in the presence of the elastic impurity scattering. The higher angular harmonics of the distribution function provide much smaller contributions to the net function $f$, due to much faster temporal relaxation. These are neglected in the eq.\ref{main}. The right side of the equation describes the inelastic relaxation of the distribution function toward the thermal equilibrium expressed by Fermi-Dirac function $f_T(\epsilon)$. The inelastic relaxation is taken in, so-called, $\tau$ approximation of the inelastic collision integral. Validity of the approximation is supported theoretically in the high temperature limit $kT \gg \hbar \omega_c$ \cite{dmitriev2005}. Below, in the numerical calculations of eq.\ref{main} we consider the inelastic scattering rate $1/\tau_{in}$ to be a constant independent on the electric field $E$ and the electron energy $\epsilon$. 

Good agreement is found between the experiment and the numerical calculations for a broad range of temperatures $kT>\Gamma$ and magnetic fields. 
At small magnetic fields the conjecture of the independence of the inelastic time  $\tau_{in}$ on the electric field $E$ is supported by direct evaluation of the variation (broadening) of the distribution function, which is found to be small at the $dc$ biases used in the experiment. The small variation provides a negligibly small correction to the inelastic collision integral and to the inelastic scattering rate. Moreover at $kT \ge \Gamma$ the energy space available for inelastic scattering of an electron inside Landau sub-band contains, in fact, all levels of the sub-band. This may provide the weak dependence of the inelastic electron scattering on the energy $\epsilon$ inside the Landau level.

At a strong magnetic field, at which Landau levels are well separated, we have found a scaling behavior of the nonlinear resistance (see fig.\ref{tau_vsT_s2},\ref{tau_vsT_s1}). In this regime the experiment and the theory demonstrate a remarkable correspondence even at a strong variation of the nonlinear resistance. This behavior is unexpected since the strong variation of the resistance implies a substantial deviation of the electron distribution function from the equilibrium and, therefore, an apparent inapplicability of the $\tau$ approximation with the constant $\tau_{in}$. Below we provide arguments, which shed a light on this interesting phenomenon.
    
At a strong magnetic field, at which Landau levels are well separated, the spectral diffusion between Landau levels is absent due to the lack of the available electron states ($\nu=0$). In this regime the total broadening of the distribution function is absent and, therefore, the total number of Landau levels participating in the spectral diffusion is fixed. There is, however, a spectral diffusion inside Landau levels, generating local spectral flows. Since the spectral diffusion conserves the total number of particles and since there is no electron transport between Landau levels, the total number of electrons inside any Landau level is preserved and equal to the thermal equilibrium value despite considerable deviations of the electron distribution function from the thermal equilibrium inside the level. It is clear that in this condition the total number of empty states in each Landau level is also fixed and equal to the value at the thermal equilibrium (at zero $dc$ bias). Thus for the isolated Landau levels the averaged spectral distribution of electron states, which are available for the inelastic scattering of an electron, is independent on the applied electric field. This may provide the significant stability of the inelastic relaxation rate with respect to the $dc$ bias. These arguments are valid, when the electron distribution inside a Landau level is not changing substantially with the electron energy. This regime holds at relatively high temperature: $kT >\Gamma$. 

At low temperatures $kT< \Gamma$ the only one Landau level is involved in electron transport and at the thermal equilibrium the electron distribution changes strongly inside the level. An application of a $dc$ bias changes appreciably the distribution of electrons.  At $kT< \Gamma$ the numerical calculations done in the $\tau$ approximation deviate substantially from the experiment (see fig.\ref{B0784T}c), indicating a limited applicability of the approximation at the low temperatures.  

The numerical calculations are done in several steps. The goal of the first step is to find the density of electron states $\nu(\epsilon)$ from a comparison with the experiment.  The density of states $\nu(\epsilon)$ of the 2D electrons can be approximated by different theoretical expressions \cite{uemura,ando,raikh,xie1990,endo2008}. We have found that the numerical results for the temperature dependence of the inelastic scattering rate are robust with respect to particular choice of the expressions for the density of states (see below). Most of the numerical results, presented in the paper, are obtained using a Gaussian form of the DOS \cite{raikh}:

\begin{equation}
\nu (\epsilon)=  \nu_0 \sqrt{\omega_c \tau_q}\sum \limits_n exp\left(-\frac{(\epsilon -n\omega _c)^2 }{\omega _{c}/\pi \tau _q} \right), 
\label{dos}
\end{equation}
where the $\tau_q$ is the quantum scattering time. To find the DOS we compare normalized longitudinal resistance $R_{xx}/R_0$ with the numerical evaluation of the normalized longitudinal conductivity $\sigma_{nl}/ \sigma_D$ obtained from eq.\ref{sigma_nl} with thermal equilibrium distribution function $f_T(\epsilon)$. The $R_0$ is the resistance of the sample in zero magnetic field. In the leading approximation and at classically strong magnetic field ($\omega_c \tau_{tr} \gg 1$) the two ratios equal to each other: $ R_{xx}/R_0=\sigma_{nl}/ \sigma_D$. From the comparison we have obtained the quantum scattering time $\tau_q$ and, therefore, have approximated the density of electron states in eq.\ref{dos}. Comparable values of quantum scattering time have been obtained using other methods, in particular, from analysis of magnitude of the quantum oscillations \cite{ando}.    

In the second step we use the DOS to numerically calculate the distribution function $f(\epsilon)$ using eq.\ref{main} in the limit $t \gg \tau_{in}$. In this limit the distribution function reaches a stationary state corresponding to the $dc$ response. The distribution function is calculated at different values of the electric field $E$. 

In the third step the normalized nonlinear conductivity $\sigma_{nl}/ \sigma_D$ is calculated using eq.\ref{sigma_nl} for different electric field. The results are compared with the normalized resistance $R_{xx}/R_0$. The inelastic scattering time $\tau_{in}$ is found from the best fit between dependencies of the normalized resistance $R_{xx}/R_0$ and the calculated normalized conductivity $\sigma_{nl}/ \sigma_D$ on the $dc$ bias.

\begin{figure}[tbp]
\vbox{\vspace{0 in}} \hbox{\hspace{+0.1in}} \epsfxsize 3.4 in 
\vskip -0.5in %
\epsfbox{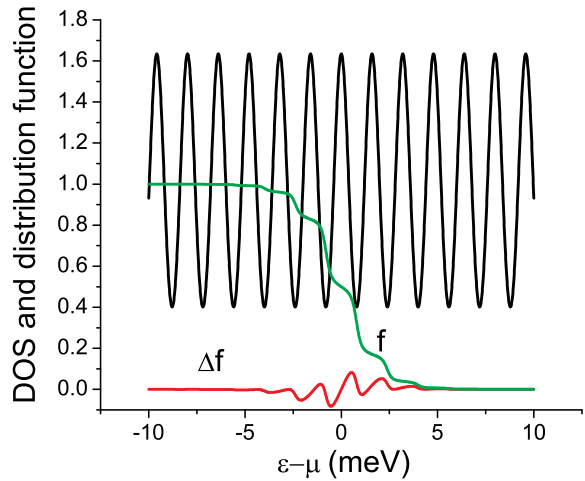} \vskip 0.5in
\caption{(color online) Normalized density of states $\tilde{\nu}$, distribution function $f$ and non-equilibrium part of the distribution function $\Delta f= f- f_T$ are shown as function of electron energy. The distribution function $f$ is obtained by numerical evaluation of eq. \ref{main}, using  physical parameters typical for experiments presented below: $I_{dc}$=377 ($\mu A$); $\tau_{in}$=0.55 (ns); $\tau_q$=1.1 (ps); B=0.924 (T) and T=10.7 (K)  }
\label{example}
\end{figure}

In accordance with eq.\ref{main} the spectral diffusion generates an electron spectral flow $J_\epsilon$ from low energy regions (occupied levels) to high energies (empty levels). The spectral flow is proportional to the coefficient of the spectral diffusion $D_\epsilon$ and to the gradient of the distribution function $\partial f/\partial \epsilon$: $J_\epsilon =D(\epsilon) \cdot \partial f/\partial \epsilon$. In a stationary state the spectral electron flow $J_\epsilon$ is constant. As a result, the gradient of the distribution function $\partial f/\partial \epsilon$ is strong in the regions of weak spectral diffusion (between Landau levels) and is small in the regions with strong spectral diffusion (centers of the Landau levels). It is important to realize that a $weak$ inelastic scattering cannot change significantly the robust dynamic flow in the energy space and, therefore, the behavior of the distribution function. This corresponds to our numerical calculations. Fig.\ref{example} demonstrates the density of states, distribution function and non-equilibrium part of the function induced by $dc$ current $I_{dc}$. Indeed the gradient of the distribution function is considerably suppressed inside Landau levels. This is due to both the fast spectral diffusion inside Landau levels and the slow diffusion between them.  Such non-equilibrium distribution function can not be described by a temperature \cite{romero2008warming}. In accordance with eq.\ref{sigma_nl} the small gradient of the distribution function inside conducting Landau levels makes the net value of the nonlinear longitudinal conductivity (resistivity) to be significantly smaller than the linear, unbiased value.  Below we present the detailed comparison between the experiments and the numerical calculations.  

\begin{figure}[tbp]
\vbox{\vspace{0 in}} \hbox{\hspace{+0.1in}} \epsfxsize 3.4 in 
\vskip -0.5in %
\epsfbox{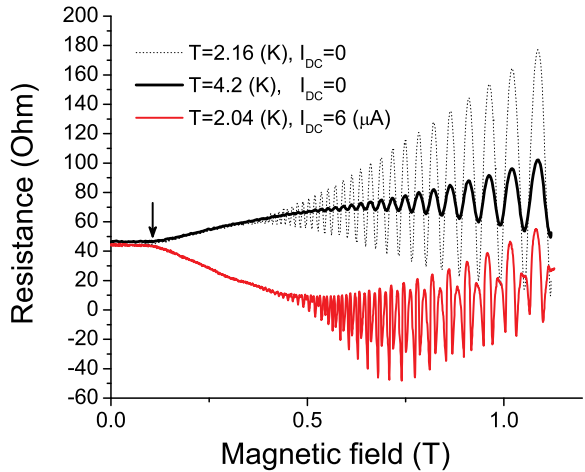} \vskip 0.5in
\caption{ (Color online), Dependencies of the longitudinal resistance $r_{xx}$ on magnetic field at different temperatures with no $dc$ bias (black solid and dotted lines) and with applied $dc$ bias $I_{dc}=$6 ($\mu$A) at T=2.04 K (grey solid line (red online)). Arrow indicates magnetic field B=0.1 T above which the electron spectrum is modulated due to quantization of electron motion: Landau levels.}
\label{SdH}
\end{figure}

\section{Results and Discussion}

Fig.\ref{SdH} demonstrates dependencies of the longitudinal resistance of 
two dimensional electrons on the magnetic field in sample N2. Two upper curves present dependencies obtained at different temperatures T=2.16K (dotted curve) and T=4.2K (solid curve) at zero $dc$ bias. At small magnetic fields $B<$0.1T the magnetoresistance demonstrates the classical independence on the magnetic field \cite{ziman}. At $B>$0.1T the electron spectrum is quantized and at temperature $T=$0.3K the resistance demonstrates quantum oscillations (not shown). An arrow marks the magnetic field $B=$0.1T above which the electron spectrum is modulated due to the quantization of the electron motion in magnetic fields.   

At magnetic fields $B<0.3$T the two traces at T=2.16K and at T=4.2K are almost identical, indicating a very weak temperature dependence of the resistance ($dr_{xx}/dT>0$). At stronger magnetic fields the quantum oscillations (Shubnikov de Haas, SdH) are observed. The oscillations are result of Landau quantization of the electron spectrum in the magnetic fields. At thermal equilibrium the amplitude $A$ of the oscillations follows from eq.\ref{sigma_nl} and eq.\ref{sigma_dc} with the Fermi-Dirac distribution function: $A \sim X_T/sinh(X_T)$, $X_T=2\pi^2kT/\hbar \omega_c$ \cite{shoenberg1984,ando}. At small magnetic fields $\hbar \omega_c \ll kT $ the amplitude of the SdH oscillations is small due to an effective averaging of the conductivity oscillations $\sigma(\epsilon)$ (see eq.\ref{sigma_dc}) over the temperature interval $kT$ in eq.\ref{sigma_nl}. Fig.\ref{SdH} shows that the increase of the temperature reduces the magnitude of the oscillations symmetrically toward a background, which is an averaged value between maximums and minimums of the oscillations.

A different behavior of the resistance is found in the response to the $dc$ bias \cite{romero2008warming}. In fig.3 the lower curve presents a typical dependence of the differential resistance on magnetic field at a finite $dc$ bias. At $B>0.1T$, at which the Landau quantization appears, the resistance shows a considerable decrease with the $dc$ bias ($dr_{xx}/dI <0$). The decrease of the resistance cannot be explained by a temperature increase due to the $dc$ heating. The temperature increase raises the resistance ($dr_{xx}/dT>0$). Moreover the quantum oscillations at the finite $dc$ bias do not have the canonical shape, corresponding to the two upper curves at zero $dc$ bias. Instead a strong increase of higher harmonics of the oscillations is obvious. The enhancement of the higher harmonic content is in apparent contradiction with the description of the $dc$ biased electrons by an elevated temperature $T_e$: high temperature reduces exponentially the higher harmonic content of the oscillations \cite{shoenberg1984,ando,romero2008warming}. 

Below we show that the strong decrease of the resistance with the $dc$ bias is result of the non-uniform spectral diffusion of 2D electrons through Landau levels. We consider in detail two regimes. One regime corresponds to small magnetic fields, at which Landau levels are overlapped and the temperature is higher than the level separation: $kT \gg \hbar \omega_c$. In this regime the quantum oscillations are absent and the resistance depends weakly on the temperature. At the small magnetic fields the spectral diffusion equation is solved both numerically and analytically\cite{dmitriev2005}. Another regime corresponds to high magnetic fields at which the Landau levels are separated: $\hbar \omega_c > \Gamma$.  For sample N2 the first regime corresponds to $B < 0.2$T  whereas the second regime is at $B>0.7$T (see fig. \ref{SdH}).

\subsection {Small magnetic fields} 

At small magnetic fields the separation between Landau levels $\hbar \omega_c$ is less than the effective width of the levels $\Gamma=\hbar/\tau_q $. At low temperatures the width $\Gamma$ is predominantly determined by the elastic impurity scattering of the 2D electrons. At small magnetic fields the density of states $\nu(\epsilon)$ is weakly oscillating with the energy $\epsilon$, making the spectral diffusion to also be a weakly modulated function of the energy. We consider a regime of high temperatures: $kT \gg \hbar \omega_c$. In this regime the quantum oscillations are absent and the resistance increases weakly with the temperature $T$.

\begin{figure}[tbp]
\vbox{\vspace{0 in}} \hbox{\hspace{+0.1in}} \epsfxsize 3.4 in 
\vskip -0.5in %
\epsfbox{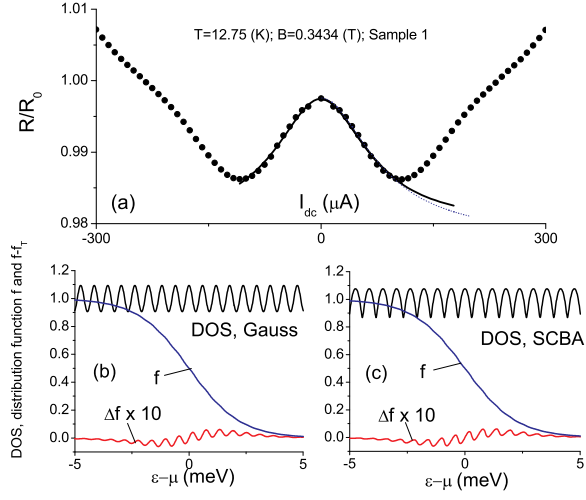} \vskip 0.5in
\caption{ (Color online), (a) Dependence of normalized longitudinal resistance $R_{xx}/(R_0=37.75 \Omega)$ on electric current . Symbols are experimental data points.  Solid lines present analytical results (eq.\ref{analytic}) and  numerical evaluation of the normalized resistance at $\gamma=0.9931$, $\tau_q=1.138$ (ps) and $\tau_{in}=$23.65 (ps) for the gaussian form of the DOS. Thin dotted line is the numerical evaluation of the resistance, using the SCBA density of states  with $\gamma=0.9931$, $\tau_q=1.132$ (ps) and $\tau_{in}=$21.4 (ps);  (b) density of states, electron distribution function $f$ and the non-equilibrium part of the function $\Delta f=f-f_T$ at $dc$ bias $I_{dc}=177.6 \mu A$, (Gaussian DOS) ; (c) density of states, electron distribution function $f$ and the non-equilibrium part of the function $\Delta f=f-f_T$ at $dc$ bias $I_{dc}=192.5 \mu A$, (SCBA DOS); T=12.75 (K), B=0.3434 (T), sample N1.}
\label{B03434T}
\end{figure}

Fig.\ref{B03434T}(a) shows the dependence of normalized resistance $R/R_0$ of the sample N1 on electric current at a small magnetic field $B=$0.343 (T) and temperature $T=$12.75 (K). The  parameter $R_0$ is the resistance at zero magnetic field. At small $dc$ biases the normalized resistance decreases with the electric current. We consider the decrease as a result of the non-uniform spectral diffusion of 2D electrons. At higher biases the resistance increases with the electric current due to other mechanisms of the nonlinearity \cite{dmitriev2007, glazman2007}. In accordance with the theory \cite{dmitriev2005} the decrease of the resistivity obeys the following relation:

\begin{equation}
\sigma_{xx}/\sigma_D=\gamma+2\delta^2[1-\frac{4Q_{dc}}{1+Q_{dc}}],  
\label{analytic}
\end{equation}
where $\gamma=1$, $\delta=exp(-\pi/\omega_c \tau_q)$ is the Dingle factor. The parameter $Q_{dc}$ takes into account the electric field $E$ ( Hall electric field \cite{hall}):    
\begin{equation}
Q_{dc}=\frac{2\tau_{in}}{\tau_{tr}}(\frac{eE
v_F}{\omega_c})^2(\frac{\pi}{\hbar \omega_c})^2. 
\label{qfactor}
\end{equation}

To compare with the experiment we have used the Dingle factor $\delta$($\tau_q$) and the inelastic scattering time $\tau_{in}$ as fitting parameters. We also have varied parameter $\gamma$ to take into account possible memory effects \cite{vavilov2004,mirlin1999} and other deviations from the Drude magnetoconductivity \cite{shklov}, which are ignored at $\gamma=1$. A solid line presents the theoretical dependence (see eq.\ref{analytic}) of the normalized resistivity at $\gamma=0.9931$, $\tau_q=1.138$ (ps) and $\tau_{in}=$23.65 (ps). Another solid line, which is indistinguishable from the analytical result, presents the numerical evaluation of the normalized resistivity, using eq.\ref{main} with the same fitting parameters $\gamma=0.9931$, $\tau_q=1.138$ (ps) and $\tau_{in}=$23.65 (ps) and the Gaussian form of the DOS \cite{raikh}.  A thin dotted line in fig.\ref{B03434T}(a) demonstrates the numerical evaluation of the resistance, using the SCBA density of states  with $\gamma=0.9931$, $\tau_q=1.132$ (ps) and $\tau_{in}=$21.4 (ps). The density of states, electron distribution function $f$ and the non-equilibrium part of the function $\Delta f=f-f_T$ are shown in fig.\ref{B03434T}(b) (Gaussian DOS) and \ref{B03434T}(c) (SCBA DOS).  Fig.\ref{B03434T}(a) demonstrates good agreement between the experiment and the theory at small $dc$ biases.     

\begin{figure}[tbp]
\vbox{\vspace{0 in}} \hbox{\hspace{+0.1in}} \epsfxsize 3.2 in 
\vskip -0.1in %
\epsfbox{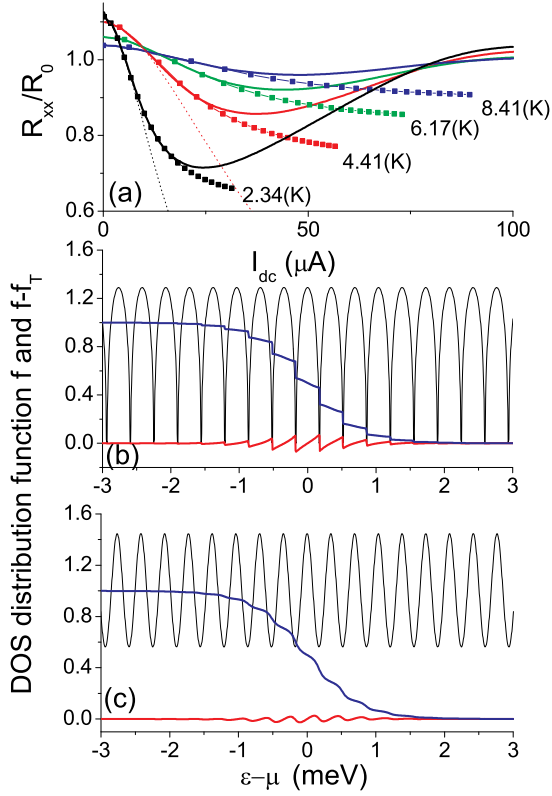} \vskip 0.0in
\caption{ (Color online), (a) Dependence of normalized longitudinal resistance $R_{xx}/R_0$ on electric current at different temperatures as labeled. Solid lines are experimental curves. Symbols present result of numerical calculations of the resistance, using Gaussian DOS (eq.\ref{dos}) with $\gamma=1$ and $\tau_q$ and $\tau_{in}$ presented in fig.\ref{B05T}(a); dotted lines demonstrate numerical evaluation of the $R/R_0$ using SCBA DOS with $\gamma=1$ and $\tau_q$ and $\tau_{in}$ presented in fig.\ref{B05T}(a).   (b) Dependencies of normalized SCBA density of states $\tilde {\nu}(\epsilon)=\nu(\epsilon)/\nu_0$,  electron distribution function $f$ and non-equilibrium part of the function $\Delta f$ on electron energy $\epsilon$ counted with respect to Fermi energy $\mu$. Distribution function a is solution of eq.\ref{main} using SCBA DOS with $\tau_q$=3.8 (ps), temperature T=4.41 (K) and electric current $I_{dc}$=50.6 ($\mu$A). (c) Dependencies of normalized Gaussian density of states $\tilde {\nu}(\epsilon)=\nu(\epsilon)/\nu_0$,  electron distribution function $f$ and non-equilibrium part of the function $\Delta f$ on electron energy $\epsilon$. The distribution function is a solution of.(\ref{main} using the Gaussian DOS with $\tau_q$=3.96 (ps), temperature T=4.41 (K) and electric current $I_{dc}$=56.4 ($\mu$A);  $R_0(2.34K)=44.6 (\Omega)$, $R_0(4.41K)= 46.36 (\Omega)$, $R_0(6.17K)= 49.29(\Omega)$, $R_0(8.41K)= 52.47(\Omega)$;  B=0.2 (T); sample N2.}
\label{B02T}
\end{figure}

Fig.\ref{B02T}(a) shows the dependence of the resistance of the sample N2 on the direct current at different temperatures as labeled. Solid lines present experimental dependencies. Dashed lines demonstrate results of numerical evaluation of the resistance, using eq.\ref{main} with SCBA DOS at T=2.34 (K) and T=4.41 (K). The numerical calculations demonstrate strong nonlinear suppression of the longitudinal resistance with the $dc$ bias. The result is due to drastic modulation of the SCBA density of states and, therefore, spectral diffusion with the energy. 

The SCBA DOS, distribution function and the non-equilibrium part of the function are presented in the fig.\ref{B02T}(b) at temperature T=4.41 (K). The DOS demonstrates sharp drops to almost zero values between Landau levels. Such strong modulation of the DOS creates significant suppression of the energy exchange between different levels facilitating the electron "warming" inside the levels \cite{romero2008warming}. The results, however, are apparently less compatible with the experiment than the one obtained with a smoother Gaussian DOS.   
\begin{figure}[tbp]
\vbox{\vspace{0 in}} \hbox{\hspace{+0.1in}} \epsfxsize 3.4 in 
\vskip -0.5in %
\epsfbox{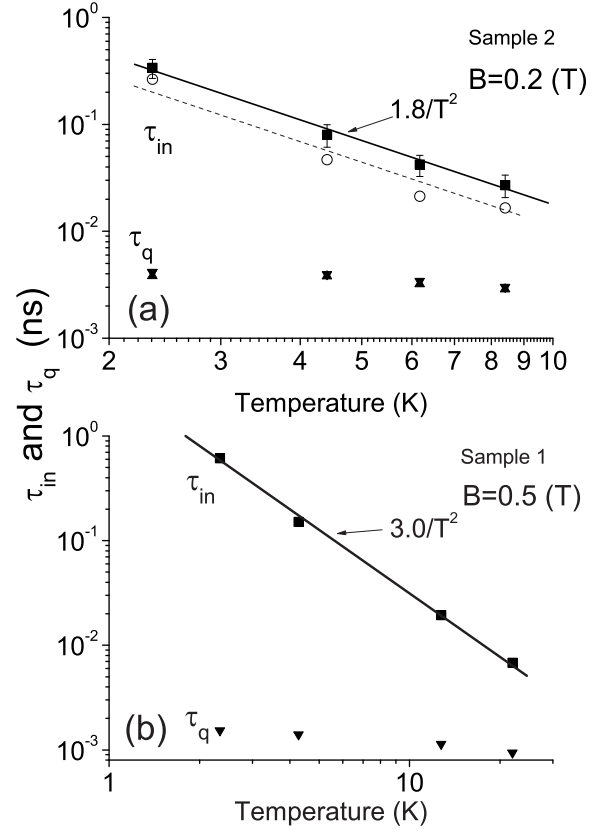} \vskip 0.5in
\caption{ Dependencies of the inelastic scattering time $\tau_{in}$ and the quantum time $\tau_q$ on temperature. (a) Filled squares show inelastic scattering time $\tau_{in}$, obtained numerically using eq.\ref{main} with Gaussian DOS; open circles present $\tau_{in}$ obtained, using eq.\ref{main} with SCBA DOS. Magnetic field B=0.2 (T). Sample N2.   
(b) Sample N1. Gaussian DOS. Magnetic field is 0.5 (T).}
\label{B05T}
\end{figure}

In fig.\ref{B02T}(a) symbols present results of the numerical evaluation of the longitudinal resistivity, using eq.\ref{main} with the Gaussian DOS and the quantum scattering times and inelastic times shown in the fig.\ref{B05T}(a). The numerical simulations demonstrate good agreement with the experiment in a considerably broader range of the $dc$ biases. Gaussian DOS is shown in the fig.\ref{B02T}(c), demonstrating moderate oscillations with energy.

The experiment and the numerical calculations correspond well to each other at small electric currents $I_{dc}$. At higher currents considerable deviations between the experiment and the theory occur. The deviations are expected. At higher currents there are additional mechanisms of the 2D electron nonlinearity \cite{yang2002,durst2003,ryzhii1970,anderson,shi,liu2005,dietel2005,inarreaPRB2005,vavilov2004,dmitriev2005,alicea2005,volkov2007,glazman2007,dmitriev2007}, which are not taken into account in eq.\ref{main}. These nonlinearities are beyond the scope of the present paper. Moreover an additional contribution to the deviations may occur due to the conjecture of the constant inelastic relaxation rate $1/\tau_{in}$ in eq.\ref{main}. At very small $dc$ biases, at which the electron distribution is near the thermal equilibrium, the variation of the inelastic rate with the $dc$ bias is also small since the phase space available for the inelastic scattering of an electron is nearly the same as at the equilibrium. At stronger $dc$ biases the distribution function is broader and the inelastic scattering rate can be considerably stronger. 

To estimate the broadening of the distribution function at small magnetic fields, at which the spectrum is weakly modulated, we approximate the distribution function by an elevated temperature $T_e$. At a stationary condition an increase of the Joule heat: $dP=d(J^2 \cdot \rho)$ is balanced by an increase of the heat dissipation: $dE/ \tau_r(T_e)=c(T_e)dT / \tau_r(T_e)$, where $c(T_e)=c_0T_e$ is the electron heat capacity, $\tau_r$ is a time of the relaxation of the total electron energy, $J$ is current density and $\rho$ is electron resistivity per square. In our case the time $\tau_r$ is controlled by the electron-phonon scattering, since the electron-electron scattering cannot stabilize the global broadening of the distribution function. For the estimation of the broadening we use $\tau_r=\tau_{e-ph}/T^3$ with $\tau_{e-ph}=20$ (ns/K$^3$) \cite{pepper,sergeev}.  An integration of both sides of the balanced equation yields:  $T_e^5-T_L^5=5\tau_{e-ph}J^2 \rho/c_0$. At the lattice temperature $T_L$=2.34 (K) the temperature increase $\Delta T_e=T_e-T_L=$0.14 (K) is found at $I_{dc}$=9 ($\mu$A). $\Delta T_e$=0.34 (K) is at $I_{dc}$=17 ($\mu$A), at which a deviation between the solution of eq.\ref{main} with a constant $\tau_{in}$ and the experiment is evident. Thus the estimation indicates that the deviation between the experiment and the theory at high $dc$ biases can be also related to the variation of the inelastic scattering time $\tau_{in}$ with the $dc$ bias. Similar results are found for sample N1.

To obtain agreement between the experimental and numerical dependencies in fig.\ref{B02T}a we have used the constant inelastic scattering time $\tau_{in}$  as a fitting parameter.  The temperature dependence of the time $\tau_{in}$, obtained from fitting at different temperatures, is shown in fig.\ref{B05T} for two samples. For sample N2 (fig.\ref{B05T}(a) black squares) the inelastic time follows the dependence $\tau_{in}=1.8 (\pm 0.3)/T^{2(\pm 0.15)}$ (ns). The time is obtained using Gaussian DOS shown in fig.\ref{B02T}(c). Open circles in fig.\ref{B05T}(a) present the inelastic time $\tau_{in}$, obtained using the SCBA DOS shown in fig.\ref{B02T}(b). The SCBA DOS results in consistently shorter inelastic times than the Gaussian DOS does, but with essentially the same temperature dependence.  This holds for other magnetic fields and temperatures. Taking into account the better overall agreement with the experiment obtained for numerical simulations with the Gaussian DOS, from now on we will only show numerical results for this density of states.      

Similar temperature dependence of the inelastic scattering time $\tau_{in}$ is found for the sample N1 with a higher electron density and considerably shorter quantum scattering time $\tau_q$.  The dependence is shown in fig.\ref{B05T}(b).  The dependence is obtained at magnetic field B=0.5 (T) and corresponds to the Gaussian DOS, which is similar to the one presented in fig.\ref{B02T}(c). The quantum scattering times $\tau_q$ in both samples are also shown for comparison and completeness in the figure. The time $\tau_q$ is much shorter the inelastic scattering time $\tau_{in}$. The quantum scattering time has weak temperature dependence. 

In accordance with the theory the temperature dependence of the inelastic time $\tau_{in} \sim T^{-2}$ indicates the dominant contribution of the electron-electron scattering into the inelastic relaxation of the distribution function. 
We have compared the experimental results with theoretical calculations of the inelastic relaxation due to electron-electron interaction  \cite{chaplik1971,quinn1982,dmitriev2005}. For the parameters corresponding to fig.\ref{B05T} the theoretical values of the inelastic time are found to be:  $\tau_{in}^{theor}=1.2/T^2$ (ns) for sample N2 (fig.\ref{B05T}(a)) and $\tau_{in}^{theor}=2.5/T^2$ (ns) for sample N1 (fig.\ref{B05T}(b)). The theoretical values are in good agreement with the experiment.  A longer inelastic relaxation, found in the experiments, could be a result of an additional screening by X-electrons in our samples \cite{fried1996}. The screening is not taken into account in the comparison.  Fig.\ref{B05T} demonstrates a longer inelastic time for sample N1 with a higher electron density in agreement with the theory \cite{chaplik1971,quinn1982,dmitriev2005}.

When considering the spectral diffusion of electrons in crossed electric and small magnetic fields at high temperatures, the results presented in this section demonstrate good quantitative agreement between the experiments and the theory. The numerical and analytical evaluation of the distribution function shows significant deviations of the electron distribution function from the Fermi-Dirac form leading to the nonlinear transport. At these conditions the rate of the inelastic relaxation of the non-equilibrium distribution function is found to be proportional to the square of the temperature: $1/\tau_{in} \sim T^2$.

\subsection {High magnetic fields}  

\begin{figure}[tbp]
\vbox{\vspace{0 in}} \hbox{\hspace{+0.1in}} \epsfxsize 3.4 in 
\vskip -0.5in %
\epsfbox{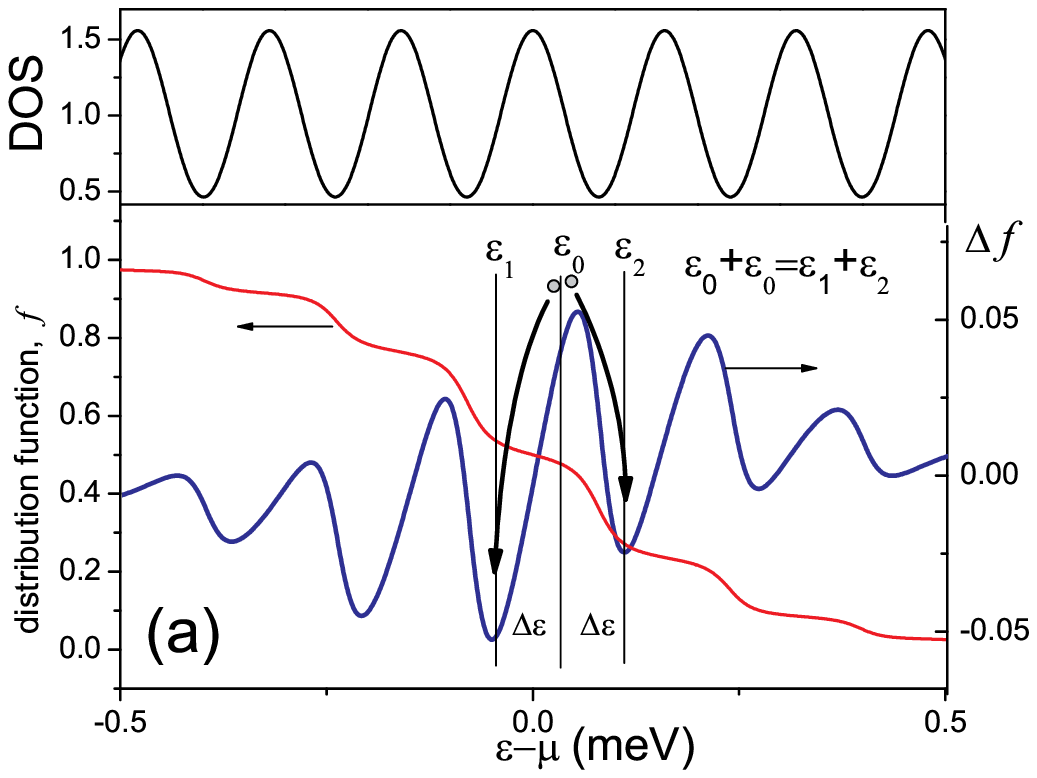} \vskip 0.5in
\vbox{\vspace{0 in}} \hbox{\hspace{+0.1in}} \epsfxsize 3.4 in 
\vskip -1in %
\epsfbox{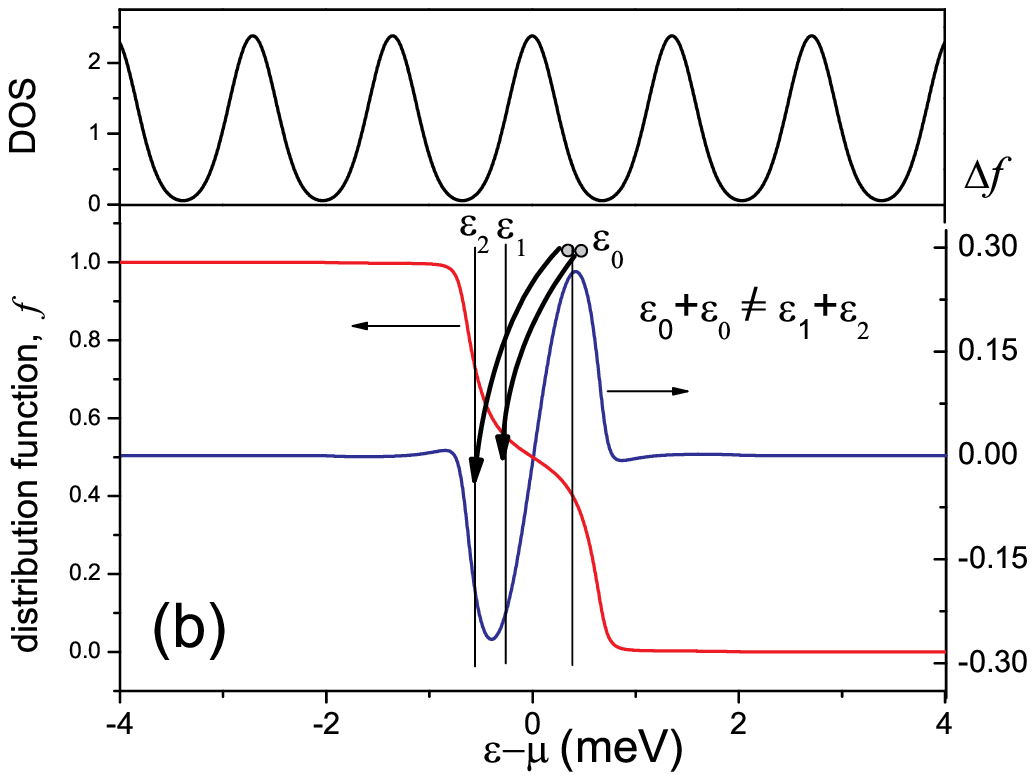} \vskip 0.5in
\caption{ (color online) (a) Relaxation of the non-equilibrium part of the distribution function $\Delta f$ by an electron-electron scattering at small magnetic fields and/or high temperatures. Two electrons near maximum of $\Delta f$ at energy $\epsilon_0$ scatter into nearest minimums at energies $\epsilon_1=\epsilon_0-\Delta \epsilon$ and $\epsilon_2 = \epsilon_0+\Delta \epsilon$. The process conserves the total electron energy $\epsilon_0+ \epsilon_0 = \epsilon_1 + \epsilon_2$ and can be accomplished by the electron-electron interaction. (b) Inelastic relaxation at high magnetic fields and/or low temperatures. The relaxation flows from overpopulated high energy levels ($\epsilon_0$) toward under-populated low energy region ($\epsilon_1, \epsilon_2 $). The relaxation flow does not conserve the total energy of 2D electron system and cannot be accomplished by $e-e$ scattering.  The electron-phonon scattering provides the relaxation.  }
\label{EEexample}
\end{figure}

At high magnetic fields the density of states and, therefore, the spectral diffusion are strongly modulated with the energy. Between completely separated Landau levels ($\Gamma \ll \hbar \omega_c$) the spectral diffusion is expected to be very weak. This may create a strong thermal isolation of the Landau levels and a stratification of the dynamic flow in the phase space in the response to the $dc$ bias. In a limiting case of a single isolated level at low temperatures the global spectral flow is absent and the slope (gradient) of the distribution function $df/d\epsilon$ is determined solely by intra-level inelastic processes. For the intra-level inelastic transitions the electron-electron interaction may not be effective, because the interaction conserves the total energy of electron system.   Fig.\ref{EEexample} demonstrates a difference between the inelastic relaxation of distribution function through several Landau levels (fig.\ref{EEexample}(a)) and the relaxation involving only  one isolated Landau level (fig.\ref{EEexample}(b)). 

The first case (fig.\ref{EEexample}(a)) corresponds to a high temperature regime: $kT \gg \hbar \omega_c$. In the first case the electron-electron interaction can effectively reduce the non-equilibrium part of the distribution function $\Delta f$ through the processes similar to the one shown in the figure. Two electrons near a maximum of the oscillating function $\Delta f$ relax into the two nearest minimums. This process reduces the non-equilibrium part of the distribution function $\Delta f$ smoothing out the oscillations. In this process the total electron energy is conserved and the relaxation can be accomplished by electron-electron scattering.   

 The second case (fig.\ref{EEexample}(b)) corresponds to low temperatures (high magnetic field) $kT < \Gamma < \hbar \omega_c$. Under these conditions the only Landau level (sub-band), located near the Fermi energy, is involved in the spectral diffusion. Lower energy levels are gapped and populated completely. They cannot participate in spectral transport due to the Pauli principle. The higher energy levels are empty, but, again, are inaccessible at low T due to the cyclotron gap.  A typical non-equilibrium part of the distribution function corresponding to this case is shown in fig.\ref{EEexample}(b).   The main flow of the relaxation to the thermal equilibrium is from overpopulated high energy levels into the under-populated low energy region of the Landau level. The relaxation flow does not conserve the total energy of electron system, and, therefore, cannot be accomplished by the electron-electron scattering. 

A possible candidate for inelastic electron relaxation is electron-phonon scattering. Electron-phonon scattering does not conserve the total electron energy and, therefore, can be the mechanism responsible for the inelastic relaxation inside the isolated Landau level at low temperatures.   Moreover, due to a stronger temperature dependence \cite{price,sergeev}, the electron-phonon scattering could be the dominant mechanism of the  relaxation at high temperature.  Below we show the interplay between different regimes of the inelastic electron relaxation, which are observed in our samples.  

\begin{figure}[tbp]
\vbox{\vspace{0 in}} \hbox{\hspace{+0.1in}} \epsfxsize 3.4 in 
\vskip -0.5in %
\epsfbox{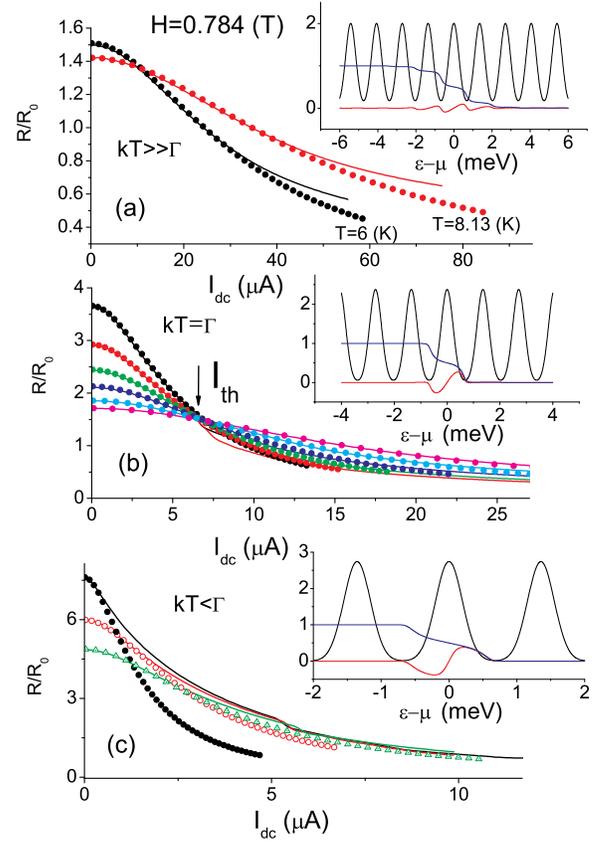} \vskip 0.5in
\caption{(color online) (a) Dependence of normalized resistance $R/R_0$ on $dc$ bias at high temperatures as labeled. $R_0(6K)=49.29 (\Omega)$, $R_0(8.13K)=52.12(\Omega)$. Insert demonstrates dependence of density of states, distribution function and non-equilibrium part of the function $\Delta f$  on energy $\epsilon$; T=8.13 (K), $I_{dc}=$58.5 ($\mu$A), $\tau_{in}$=151 (ps), $\tau_q=$1.9 (ps). (b) Dependence of normalized resistance on $dc$ bias at intermediate temperatures from top to bottom at zero bias: T=1.48($R_0=43.68(\Omega)$), 1.97($R_0 =44.33 (\Omega)$), 2.44($R_0 =44.99 (\Omega)$), 2.93($R_0=45.45 (\Omega)$), 3.52($R_0=45.89 (\Omega)$), 4.08($R_0= 46.37 (\Omega)$) (K). The electron system undergoes a transition to state with zero differential resistance at $I_{dc}> I_{th}$ and $T<$3 (K). Insert demonstrates dependence of density of states, distribution function and non-equilibrium part of the function $\Delta f$  on energy $\epsilon$; T=2.44 (K), $I_{dc}=$18.2 ($\mu$A), $\tau_{in}$=3.77 (ns), $\tau_q=$2.75 (ps). (c) Dependence of normalized resistance on $dc$ bias at low temperatures from top to bottom at zero bias: T=0.27($R_0 =42 (\Omega)$), 0.71($R_0 =42.64 (\Omega)$), 1.06($R_0 =42.99 (\Omega)$) (K). Insert demonstrates dependence of density of states, distribution function and non-equilibrium part of the function $\Delta f$  on energy $\epsilon$; T=0.71 (K), $I_{dc}=$6.67 ($\mu$A), $\tau_{in}$=17.7 (ns), $\tau_q=$3.65 (ps). Symbols are numerical calculations and solid lines are experiments. Magnetic field is 0.784 (T). Sample N2.     }
\label{B0784T}
\end{figure}  

Fig.\ref{B0784T}(a) presents dependencies of the normalized resistance of the sample N2 at $B=0.784$ (T) and at high temperatures as labeled. The magnetic field corresponds to a maximum of the SdH oscillations. At small currents the numerical simulation describes well the experiment. The insert to the figure shows the normalized density of states, distribution function $f$ and non-equilibrium part of the function $\Delta f$ at $dc$ bias 58.5 ($\mu$A). The regime corresponds to the condition $kT \gg \Gamma$. 

 Fig.\ref{B0784T}(b) presents dependencies of the normalized resistance at medium temperatures $kT \sim \Gamma$.  Again, at small currents the numerical simulation, obtained in the $\tau_{in}$ approximation of the right side of eq.\ref{main}, works well, providing  very good fit of the experiment data. At temperatures below 3 (K) a sudden deviation between the experimental data and the simulation occurs above a threshold current of $I_{th}$= 6.6 ($\mu$A). An arrow in the figure marks this current. It has been shown, that above the current $I_{th}$ the electron system undergoes a transition into the zero differential resistance state \cite{bykov2007zdr,zudov2008zdr}. In this state the differential resistance of the sample is nearly zero in a broad range of the current $I_{dc} > I_{th}$. Non-uniform, domain-like structures, propagating in real space, have been proposed to explain the origin of the electron state with zero differential resistance\cite{bykov2007zdr,vavilov2004}. Such states are beyond the regime described by the spatially uniform eq.\ref{main}. 

It is interesting that the transition to the nonlinear state with zero differential resistance happens at a normalized value of the resistance $R_{tr}=R/R_0 \approx 1.5$, which is almost independent on the temperature. Moreover at this point ($R_{tr}, I_{th}$) the nonlinear resistance demonstrates a transition from an insulating-like ($dR/dT<0$) to a metallic-like ($dR/dT>0$) behavior. These unexpected features are currently not understood and will be subject of future studies. The insert to the figure shows the normalized density of states, distribution function $f$ and non-equilibrium part of the function $\Delta f$ obtained at $dc$ bias 18.2 ($\mu$A). 

Finally fig.\ref{B0784T}(c) presents data at very low temperature $kT < \Gamma$. At this condition only one Landau level provides the electron transport. At the low temperatures the theory, used in the $\tau_{in}$ approximation, fits with the data only at very small currents. At the lowest temperature T=0.27K, numerical results deviate almost immediately from the experiment.  The comparison indicates that the approximation of the inelastic collision integral in eq.\ref{main} by a constant relaxation time $\tau_{in}$ does not work in these conditions.  At very low temperature the equilibrium distribution changes very rapidly with the energy $\epsilon$ inside the Landau level on a scale, which is much narrower than the level width $\Gamma$: $kT \ll \Gamma$. Since the inelastic processes are extremely weak at the low T, the spectral diffusion broadens easily the electron distribution to a scale comparable with the width of the level $\Gamma $ even at small $dc$ biases. This process increases significantly the phase space available for the inelastic electron scattering, enhancing the scattering rate $1/\tau_{in}$ appreciably.     Thus at $kT < \Gamma$ the inelastic scattering depends strongly on the $dc$ bias and the spectral diffusion equation (eq.\ref{main}) with a constant $\tau_{in}$ does not describe the nonlinear resistance appropriately. More work is required to evaluate quantitatively the shape of the distribution function in this regime. However we suggest that even in the regime $kT < \Gamma$ the distribution function will be qualitatively similar to the one shown in the insert to fig.\ref{B0784T}(c), which is obtained in the $\tau$ approximation. At a high $dc$ bias the function can not be described by an elevated electron temperature as it is shown in the figure (see also \cite{romero2008warming}).

\begin{figure}[tbp]
\vbox{\vspace{0 in}} \hbox{\hspace{-0.3in}} \epsfxsize 3.0 in 
\vskip -0.5in %
\epsfbox{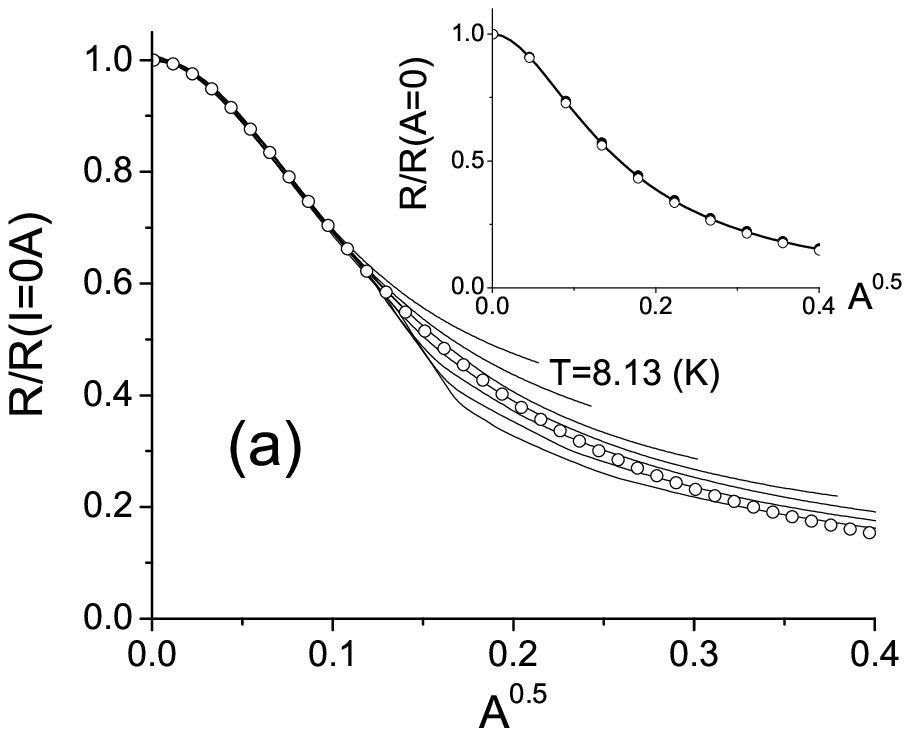} \vskip -0.5in
\vbox{\vspace{0 in}} \hbox{\hspace{-0.3in}} \epsfxsize 3.0 in
\epsfbox{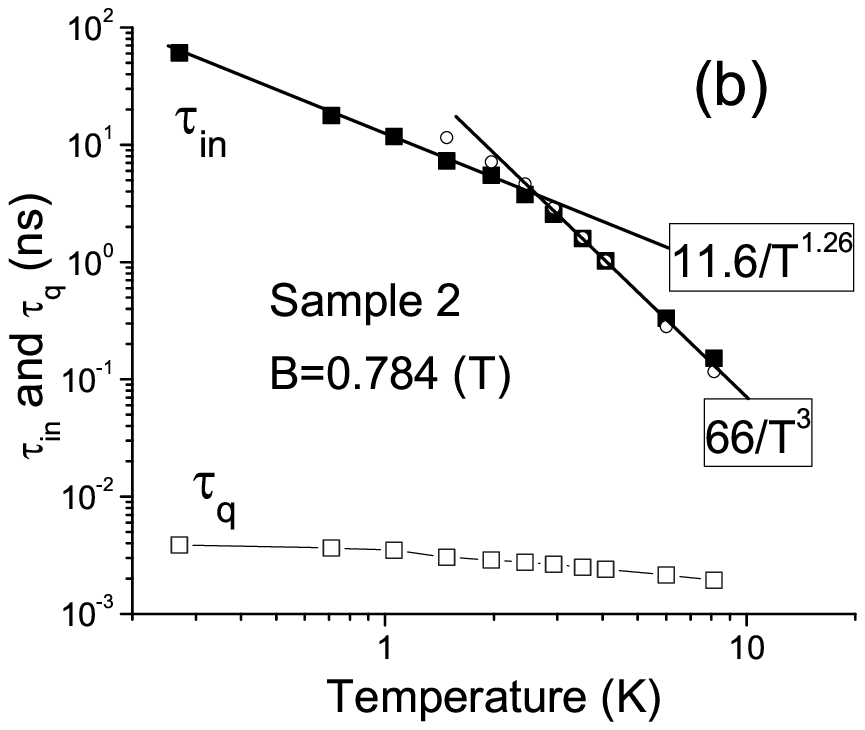} \vskip 0.5in 
\caption{ (a) Scaling of normalized resistance with parameter $A^{0.5} \sim I_{dc}$. All curves presented in fig.\ref{B0784T}(a,b) at different temperatures (1.48-8.13) (K) follow the same dependence on the parameter $A^{0.5}<0.15$ (solid curves). Open circles present results of numerical calculations of the normalized resistance, using eq.\ref{main} with $\tau_q=2.75$ (ps), $T=4.08$ (K), B=0.784 (T) and parameter $A^{1/2}=(\sigma_{dc}^D E^2 \tau_{in}/\nu_0)^{1/2}$; insert shows independence of variations of the normalized resistance with $A$ on temperature $T$. The results are obtained using eq.\ref{main} at T=3(K) -open circles, T=4.08(K) - solid curve, and T=6(K)-filled circles.
(b)Dependences of inelastic scattering time $\tau_{in}$, obtained from comparison between experiment and numerical evaluation of nonlinear resistance, using eq.\ref{main} (filled squares) and from scaling (open circles) on temperature. Open squares present temperature dependence of quantum scattering time $\tau_{q}$. Magnetic field $B$=0.784 (T). Sample N2.}
\label{tau_vsT_s2}
\end{figure}

\begin{figure}[tbp]
\vbox{\vspace{0 in}} \hbox{\hspace{-0.3in}} \epsfxsize 3.0 in 
\vskip -0.5in %
\epsfbox{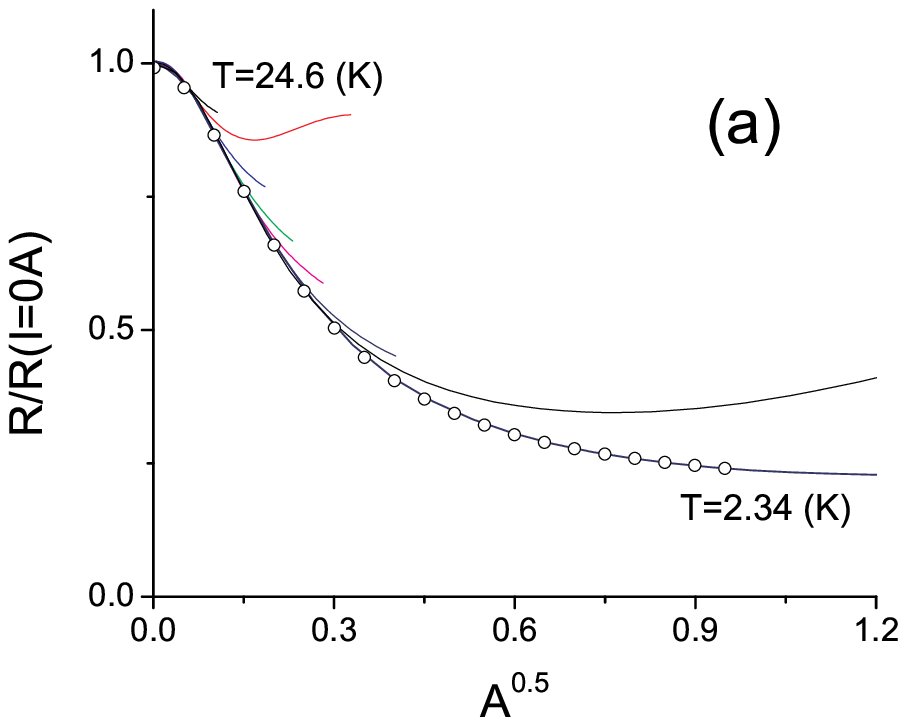} \vskip -0.5in
\vbox{\vspace{0 in}} \hbox{\hspace{-0.3in}} \epsfxsize 3.3 in
\epsfbox{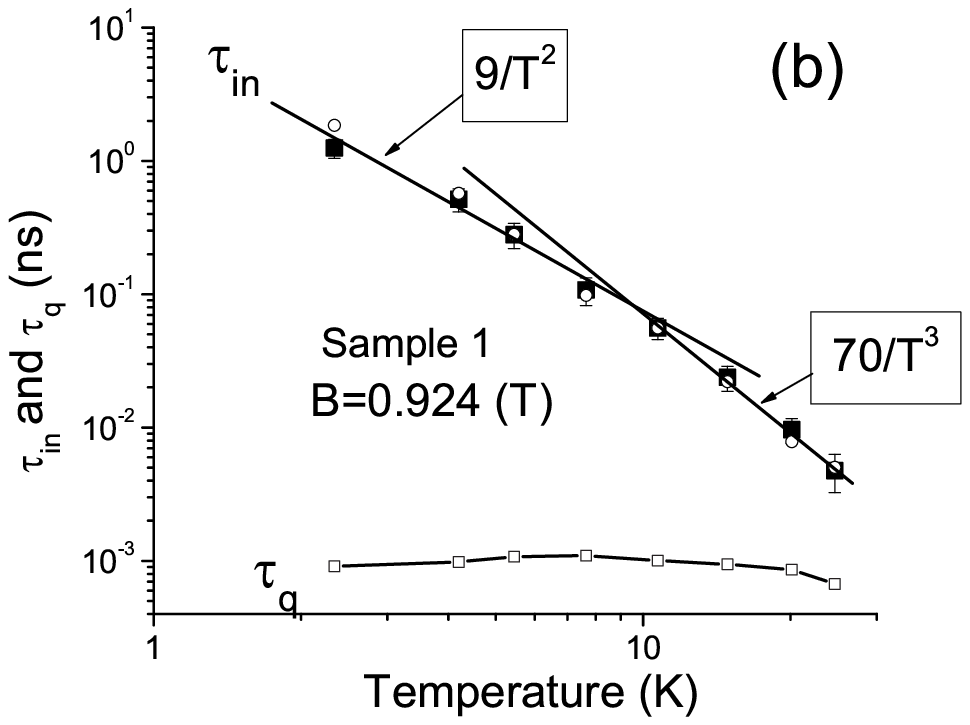} \vskip 0.5in
\caption{(a) Scaling of normalized resistance (solid curves) with parameter $A^{0.5} \sim I_{dc}$ at different temperature from bottom to top: 2.34, 4.2, 5.4, 7.6, 10.7, 14.8, 20.1, 24.6 (K). Open circles present results of numerical calculations of the normalized resistance, using eq.\ref{main} with $\tau_q=1.1$ (ps), $T=2.34$ (K), $B$=0.924 (T) and parameter $A^{1/2}=(\sigma_{dc}^D E^2 \tau_{in}/\nu_0)^{1/2}$; (b)Dependences of inelastic scattering time $\tau_{in}$, obtained from comparison between experiment and numerical evaluation of nonlinear resistance, using eq.\ref{main} (filled squares) and from scaling (open circles) on temperature. Open squares present temperature dependence of quantum scattering time $\tau_{q}$. Magnetic field $B$=0.924 (T). Sample N1.}
\label{tau_vsT_s1}
\end{figure}

Additional analysis of the curves at the high magnetic fields reveals an interesting scaling behavior of the nonlinear resistance. Applying two linear transformations ($y^{'}=K_y \cdot y$ and $x^{'}=K_x \cdot x$) along y and x-axes one can collapse all dependencies at different temperatures presented in fig.\ref{B0784T}(a,b) on a single curve. Fig.\ref{tau_vsT_s2}(a) shows the result. The y-transformation normalizes the resistance at zero bias to unity: $R(I)=R(I)/R(I=0)$.  The linear x-transformation, applied along the x-axes, provides the final result. Solid curves are experimental dependencies measured in temperature interval (1.48-8.13) (K). Open circles show a result of numerical calculations of the nonlinear resistance obtained using eq.\ref{main} with the equilibrium electron distribution at $T=$4.08 (K) and $\tau_q=2.75$ (ps). The same scaling is found for sample N1 in a broader range of temperatures. The result is shown in fig.\ref{tau_vsT_s1}(a).  All dependencies are plotted versus a parameter $A^{1/2}=(\sigma_{dc}^D E^2 \tau_{in}/\nu_0)^{1/2} \sim I_{dc}$. At a fixed density of states $\nu(\epsilon)$ the variable $A \sim E^2 \tau_{in}$ is the main parameter, which determines the deviation of the electron distribution $f$ from the thermal equilibrium $f_T$ in eq.\ref{main}. 

Fig.\ref{tau_vsT_s2}(a) demonstrates a good scaling and a remarkable correspondence with numerical results obtained at $A^{1/2}<0.15$, using eq.\ref{main} with a fixed $\tau_{in}$. The correspondence between the experiment and the theory is even more impressive for a curve at the lowest temperature (T=2.34 (K)) presented in fig.\ref{tau_vsT_s1}(a). Almost perfect agreement between the experiment at T=2.34 (K) and the theory is found at substantially stronger $dc$ biases ($A \sim 1$).  The scaling of the nonlinear resistance and the excellent agreement with the theory indicates strongly the presence of the spectral diffusion with a constant rate of the inelastic relaxation $1/\tau_{in}$. 

We suggest that the scaling is a result of a specific nonlinear regime, which occurs for separated Landau levels. As we have already mentioned in the section "Theory and Numerical Simulations", the spectral diffusion between well-separated Landau levels is absent. In this regime there is no global broadening of the distribution function. Moreover inside each of the Landau levels the local spectral flow preserves the number of electrons and, therefore, the number of the empty states. Thus the stratified spectral diffusion keeps the spectral distribution of the available phase space (averaged over each Landau level), to be fixed and the same as the one at the thermal equilibrium ($E=0$). The invariance of the phase space available for inelastic processes could provide the independence of the inelastic scattering time $\tau_{in}$ on the $dc$ bias fixing the time at the thermal equilibrium value: $ \tau_{in}(E)=\tau_{in}(E=0)$. The constant inelastic scattering rate makes the evolution of the electron distribution and the nonlinear resistance to be universal in a broad range of the $dc$ biases. 

The scaling reveals another interesting property of the nonlinear regime. 
Fig.\ref{tau_vsT_s2}(a)  shows that variations of the normalized resistance with parameter $A^{1/2}<$0.15 is the same at different temperatures and, therefore, does not depend on the initial, equilibrium distribution $f_T$ of 2D electrons in eq.\ref{main}. The equilibrium distribution $f_T$ is substantially different in the temperature interval, in which the scaled dependencies have been measured: (1.4 - 8.13) (K). We suggest that the independence of the nonlinear resistance on the $f_T$ is also a result of the absence of the $dc$ bias induced spectral flows between Landau levels. Without the inter-level spectral flow the levels are, in essence, independent from each other and, therefore, absorb the energy from electric field independently. The absorption inside each Landau level is determined by the same spectral dynamics, assuming that the density of states is the same for each level.  An estimation of the nonlinear conductivity in a model of separated (independent) levels supports the suggestion \cite{vitkalov2007unpublished}.  The numerical evaluation of the nonlinear behavior of the resistance, which has been done for different temperatures, using eq.\ref{main}, demonstrates also the independence of the normalized nonlinear resistance on the temperature in this regime.  In particular, the numerical values of the normalized resistance obtained for T=3K, T=4.08K and T=6K at a fixed density of states  ($\tau_q=2.75$ (ps)) differ by less that 3\% at any $A<$0.4. This is shown in the insert to fig.\ref{tau_vsT_s2}(a).  

The scaling of the nonlinear resistance provides an easy practical access to the variation of the inelastic relaxation time with the temperature since it does not require the solution of the eq.\ref{main}. The scaling coefficient $K_x \sim E \cdot (\tau_{in}(T))^{1/2}$ takes into account the temperature variations. A comparison of the inelastic time $\tau_{in}$ obtained from the scaling (open circles) and from the direct comparison with the numerical calculation of the nonlinear resistance using eq.\ref{main} (solid squares) are presented in fig.\ref{tau_vsT_s2}(b) (sample N2) and fig.\ref{tau_vsT_s1}(b) (sampleN1). There is a good overall agreement between two approaches. A difference appears since the numerical calculation takes into account a variation of spectral dynamics with the temperature due to changes in density of states (see the time $\tau_q$ presented in the figures) and a temperature variation of the transport scattering rate.

Deviations from the scaling depend on the temperature.
Presented in fig.\ref{tau_vsT_s2}(a) and fig.\ref{tau_vsT_s1}(a) at higher temperatures experimental curves deviate up from the scaling behavior at a smaller $A$. Taking into account the strong reduction of the inelastic scattering time $\tau_{in}$ with the temperature, one can find that the deviations from the scaling occur at progressively higher $dc$ biases: $E \sim (A/\tau_{in})^{1/2}$. This indicates that corrections to the scaling due to other nonlinear mechanisms, arising at high biases \cite{yang2002, glazman2007,dmitriev2007}, decreases with the temperature increase. The later agrees with the temperature dumping of a magnitude of the $dc$ bias induced magneto-oscillations of the nonlinear resistance \cite{zudov2009} due to inter-level scattering \cite{yang2002}. At high $dc$ biases $A^{1/2}>0.15$ sample N2 demonstrate an additional abrupt deviation down from the scaling at temperatures below 3K (see fig.\ref{tau_vsT_s2}(a)). As we have mentioned at this condition a transition to the zero differential resistance state appears \cite{bykov2007zdr,zudov2008zdr}, which may break down the description of the 2D electron system by the spatially uniform spectral equation (eq.\ref{main}) \cite{bykov2007zdr}.

Below we discuss the temperature dependence of the inelastic scattering time.
Fig.\ref{tau_vsT_s2}b presents the temperature dependence of the time $\tau_{in}$ at magnetic field $B$=0.784 (T) for the sample N2.  Two temperature regimes are clearly observable. At temperatures $T>2$K the inelastic relaxation time $\tau_{in}$ is inversely proportional to $T^3$: $\tau_{in}=66(\pm 10)/T^{3 (\pm 0.15)}$ (ns). At temperatures below 2K the inelastic time depends weaker on the temperature:  $\tau_{in}=11.6 (\pm 2)/T^{1.26 \pm 0.15}$ (ns).  

The observed $T^3$ dependence of the inelastic time $\tau_{in}=66/T^3$(ns) correlates with the one obtained in Si-MOSFETs : $\tau_{in}=(10-60)/T^3$ (ns) at temperatures $1.5<T<4.2$K \cite{dolgopol1985} and with the dependence found in a GaAs/AlGaAs heterojunction: $\tau_{in}=20/T^3$(ns) at temperatures $1<T<3$K \cite{pepper}. In both papers the temperature dependence has been attributed to the electron-phonon scattering. We suggest that the temperature dependence observed at $T>$2K is also due to an electron-phonon scattering in Bloch-Gruneisen (BG) regime at which the wave vector of a typical thermal phonon $q_T=kT/\hbar s$ is smaller than the size of the Fermi circle $2k_F$: $q_T < 2k_F$. Here $s$ is sound velocity and $k_F$ is Fermi wave vector \cite{ziman}.  In our high density samples the BG regime exists at temperatures below $T_{BG} \approx 20$K, where $kT_{BG}=2k_F \cdot \hbar s$ \cite{stormer1990}. A theoretical evaluation of the inelastic electron-phonon scattering time in GaAs quantum wells due to screened piezoelectric (PZ) coupling yields: $\tau_{PZ} \approx 16/T^3$ (ns) at temperatures of few K at zero magnetic field \cite{price,karpus1996}. Deformation potential (DP) yields a comparable contribution to the electron-phonon scattering rate at $T>$4K. At a weak screening the electron-phonon scattering time is found to be  $\tau_{DP} \approx 18/T^3$(ns) \cite{sergeev} at zero magnetic field.

The $T^{-3}$ temperature dependence is found also for the sample N1 at high temperatures. Fig.\ref{tau_vsT_s1}b  presents the temperature dependence. At $T>$10K the inelastic scattering time is proportional to $1/T^3$: $\tau_{in}=70(\pm 10)/T^{3 \pm 0.2}$ (ns). The dependence is the same as the one observed in the sample N2. At lower temperatures $T<10$ (K) the inelastic relaxation time deviates consistently from the $T^{-3}$ dependence. The temperature dependence $\tau_{in}=9 (\pm 2)/T^{2(\pm 0.2)}$ provides a reasonable approximation, indicating a possible contribution of the electron-electron interaction to the inelastic relaxation rate. The same ($T^{-2}$) temperature dependence is observed at small magnetic fields for both samples but at considerably stronger relaxation rate. 
Thus the temperature dependence below 10(K) appears as an intermediate regime at which the electron-electron scattering is significant but is suppressed considerably by the quantization of the electron spectrum. At the beginning of the section we have discussed the possible reason for the reduction of the contribution of the $e-e$ scattering to the inelastic relaxation in strong magnetic fields.

\begin{figure}[tbp]
\vbox{\vspace{0 in}} \hbox{\hspace{+0.1in}} \epsfxsize 3.4 in 
\vskip -0.5in %
\epsfbox{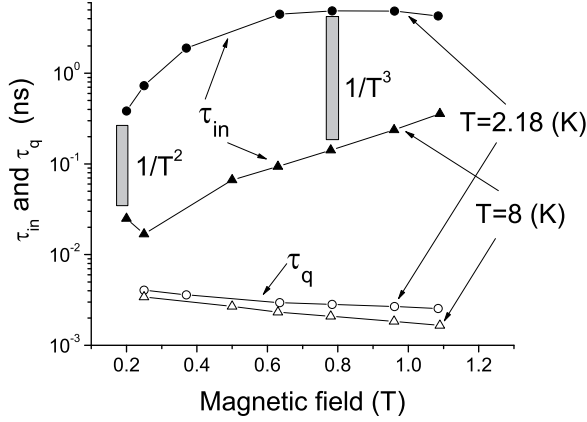} \vskip 0.5in
\caption{ Dependences of inelastic scattering time $\tau_{in}$ and quantum scattering time $\tau_{q}$ on magnetic field at two different temperatures as labeled. Two shaded areas indicate two different temperature regimes of the inelastic electron relaxation observed in the sample. Sample N2.}
\label{tau_vsB_s2}
\end{figure}

Our experiment demonstrates a correlation between modulation of the density of states, the inelastic time $\tau_{in}$ and the temperature dependence of the time. At low magnetic field $B$=0.2 (T) the density of states of the sample N2 is weakly modulated at about $\pm$40\% (see fig.\ref{B02T}). The time of inelastic relaxation equals to $1.8/T^2$ (ns) below 8K.  At the magnetic field $B=$0.784 (T) the modulation of the density of states of the sample N2 is significantly stronger approaching 95 \% of the averaged value (see fig.\ref{B0784T}(a,b)). The inelastic time equals to $66/T^3$ at $2<T<8$ (K). In magnetic field $B=$0.924 (T) the modulation of the density of states of the sample N1 is about 60\% and the inelastic time is between the two previous values: $1.8/T^2 < 9/T^2 < 66/T^3$ at $T<7$ (K).   

In accordance with the correlation one should expect a gradual reduction of the contribution of electron-electron scattering to the inelastic relaxation and an increase of the relaxation time $\tau_{in}$ with an increase of the modulation of the density of states. An increase of the magnetic field $B$ enhances the DOS modulation. Fig.\ref{tau_vsB_s2} presents the dependence of the inelastic time $\tau_{in}$ on the magnetic field for  sample N2 at two different temperatures as labeled. Magnetic field increases the relaxation time $\tau_{in}$. The temperature dependence of the inelastic relaxation rate changes from $T^2$ at low magnetic field to $T^3$ at high magnetic fields. In the figure, two rectangular shaded areas indicate the two different temperature regimes of the inelastic relaxation.  These regimes are presented in more details in fig.\ref{B05T}(a) and fig.\ref{tau_vsT_s2}(b). Similar enhancement of the relaxation time $\tau_{in}$ with the increase of the magnetic field is found for sample N1 (not shown).

\section{Conclusion}
We have studied the nonlinear response of 2D electrons placed in crossed electric and quantized magnetic fields at low temperatures. The resistance of 2D electrons decreases strongly with an increase of the electric field. The decrease of the resistance is in good quantitative agreement with theory considering the nonlinear response as a result of non-uniform spectral diffusion of 2D electrons limited by inelastic electron scattering. Comparison between the experiments and the theory has revealed different regimes of the electron inelastic relaxation.  

At low magnetic fields, at which the Landau levels are well overlapped and the spectral diffusion is weakly modulated with the electron energy, the inelastic scattering rate is found to be proportional to the square of the temperature $T^2$ in temperature interval (2-10 (K)). The dependence indicates the electron-electron scattering as the dominant mechanism of the inelastic relaxation. At high magnetic fields, at which the Landau levels are well separated, the spectral diffusion is strongly modulated and the rate of the inelastic relaxation is proportional to $T^3$. This suggests the electron-phonon scattering to be the dominant inelastic mechanism. At fixed temperature the inelastic time $\tau_{in}$ increases with the magnetic field. At very small temperatures $kT<\Gamma$ and well separated Landau levels an additional regime of the inelastic electron relaxation is identified: $1/ \tau_{in} \sim T^{1.26}$.

At the high magnetic fields the nonlinear resistance demonstrates scaling behavior in a broad range of temperatures exceeding the width of Landau levels. The scaling indicates specific regime of the $dc$ heating in electron systems with discrete electron spectrum.  A temperature cannot describe the heating. The spectral diffusion limited by the inelastic relaxation with constant rate describes remarkably well the scaling in broad range of the $dc$ biases.

\begin{acknowledgements}

S. Vitkalov thanks I. Aleiner, I. Dmitriev and A. Sergeev for valuable discussions and comments. This work was supported by National Science Foundation: DMR 0349049 and by Russian Fund for Basic Research, project No.08-02-01051 

\end{acknowledgements}


\begin{references}
\bibitem{hall} In classically strong magnetic field ($\omega_c \tau_{tr} \gg 1$) the electric field $E$ is almost perpendicular to the current $I_{dc}$. The field is determined from relation $E=I_{dc}/(R_{xy} \cdot d)$, where $R_{xy}$ is Hall resistance and $d$ is width of the sample. In the experiments the Hall resistance demonstrated negligibly small variations with the $dc$ bias.
\end{references}
\end{document}